\documentstyle[12pt,psfig]{article}

\def\ib#1#2#3{           {\it ibid. }{\bf #1} (19#2) #3}
\def\nps#1#2#3{          {\it Nucl. Phys. B (Proc. Suppl.) }
                         {\bf #1} (19#2) #3} 
\def\np#1#2#3{           {\it Nucl. Phys. }{\bf #1} (19#2) #3}
\def\pl#1#2#3{           {\it Phys. Lett. }{\bf #1} (19#2) #3}
\def\pr#1#2#3{           {\it Phys. Rev. }{\bf #1} (19#2) #3}
\def\prep#1#2#3{         {\it Phys. Rep. }{\bf #1} (19#2) #3}
\def\prl#1#2#3{          {\it Phys. Rev. Lett. }{\bf #1} (19#2) #3}
\def\rmp#1#2#3{          {\it Rev. Mod. Phys. }{\bf #1} (19#2) #3}
\def\zp#1#2#3{           {\it Zeit. fur Physik }{\bf #1} (19#2) #3}
\def\sjnp#1#2#3{         {\it Sov. J. Nucl. Phys. }{\bf #1} (19#2) #3}
\def\jetp#1#2#3{         {\it Sov. Phys. JETP }{\bf #1} (19#2) #3}
\def\jetpl#1#2#3{         {\it JETP Lett. }{\bf #1} (19#2) #3}
\def\ppnp#1#2#3{           {\it Prog. Part. Nucl. Phys. }{\bf #1} (19#2) #3}
\def\eq#1{{eq. (\ref{#1})}}
\def\ne{\hbox{$\nu_e$ }}
\begin{document}
\thispagestyle{empty}
\begin{titlepage}
\begin{center}
\hfill hep-ph/9711316\\
\hfill FTUV/97-43\\
\hfill IFIC/97-59\\
\vskip 0.3cm
\LARGE
{\bf Low-energy Anti-neutrinos from the Sun}
\end{center}
\normalsize
\vskip1cm
\begin{center}
{\bf S. Pastor}\footnote{
E-mail: sergio@flamenco.ific.uv.es},
{\bf V.B. Semikoz}
\footnote{E-mail: semikoz@izmiran.rssi.ru.
On leave from the {\sl Institute of the
Terrestrial Magnetism, the Ionosphere and Radio 
Wave Propagation of the Russian Academy of Sciences, 
IZMIRAN, Troitsk, Moscow region, 142092, Russia}.}
{\bf and J.W.F. Valle}
\footnote{E-mail valle@flamenco.ific.uv.es}\\
\end{center}
\begin{center}
\baselineskip=13pt
{\it Instituto de F\'{\i}sica Corpuscular - C.S.I.C.\\
Departament de F\'{\i}sica Te\`orica, Universitat de Val\`encia\\}
\baselineskip=12pt
{\it 46100 Burjassot, Val\`encia, SPAIN         }\\
\vglue 0.8cm
\end{center}
 
\begin{abstract}

We consider the sensitivity of future neutrino experiments in the low
energy region, such as BOREXINO or HELLAZ, to a solar $\bar{\nu}_e$
signal.  We show that, if neutrino conversions within the Sun result
in partial polarization of initial solar neutrino fluxes, then a new
opportunity arises to observe the $\bar{\nu}_e$'s and thus to probe
the Majorana nature of the neutrinos. This is achieved by comparing
the slopes of the energy dependence of the differential $\nu
e$-scattering cross section for different neutrino conversion
scenarios.  We also show how the $\nu_e \rightarrow \bar{\nu}_e$
conversions may take place for low energy solar neutrinos while being
unobservable at the Kamiokande and Super-Kamiokande experiments.

\end{abstract}
 
\vfill
 
\end{titlepage}

\vskip 1cm

\section{Introduction}

Finding a signature for the Majorana nature of neutrinos or,
equivalently, for the violation of lepton number in Nature is a
fundamental challenge in particle physics \cite{fae}. All attempts for
distinguishing Dirac from Majorana neutrinos {\em directly} in
laboratory experiments have proven to be a hopeless task, due to the
V-A character of the weak interaction, which implies that all such
effects vanish as the neutrino mass goes to zero. This applies to all
searches for processes such as neutrino-less double beta decay
\cite{bb}, as well as the proposal to search for CP violating
effects induced by the so-called Majorana CP phases \cite{2227} in
neutrino propagation \cite{1666}.  In this paper we suggest an
alternative way in which one might probe for the possibility of
$L$-violation which it is not {\em directly} induced by the presence
of a Majorana neutrino mass. Of course, Majorana masses will be
required at some level, as it must be the case due to a general
theorem \cite{BOX}, but the quantity which is directly involved is the
transition amplitude for a \ne to convert into an anti-\ne inside the
Sun. One way to achieve this is via a non-zero transition magnetic
moment of Majorana neutrinos \cite{BFD} which may be resonantly
enhanced due to the effect of matter in the Sun \cite{LAM}.

In this note we argue that the possible observation of an electron
anti-neutrino component in the solar neutrino flux at low energies
could provide an indication that neutrinos are Majorana particles.
Future real-time neutrino experiments, like HELLAZ \cite{hellaz},
BOREXINO \cite{borexino} or HERON \cite{heron}, have been proposed to
directly measure the fluxes of low-energy solar neutrinos using the
neutrino-electron scattering reaction. BOREXINO is designed to take
advantage of the characteristic shape of the electron recoil energy
spectrum from the $^7$Be neutrino line. The measurement of the $^7$Be
neutrino flux in BOREXINO will play a key r\^ole in clarifying the
solar neutrino problem and in discriminating which one of all the
proposed physical scenarios where neutrinos have non-standard
properties is the correct one.  For instance, if a very small ratio
$R_{Be}^{exp}/R_{Be}^{SSM}$ were measured, as expected in order to
reconcile the data from Homestake and Kamiokande/Super-Kamiokande
\cite{castellani}, this would point towards the small-mixing MSW 
solution. On the contrary, if this ratio were found to be larger,
there are many ways to explain the deficit, such as the large-mixing
MSW solution. The measurement of the $^7$Be neutrino line at BOREXINO
would also test the level of density fluctuations in the solar matter
as recently shown in ref. \cite{noise}. On the other hand, the HELLAZ
and HERON experiments are intended to measure the fundamental
neutrinos of the $pp$ chain.

It is well-known that if neutrinos are Majorana particles they
can not have non-zero magnetic moments. However, they can have
transition magnetic moments \cite{BFD} which may induce
chirality-flipping transitions such as
\begin{equation}
\label{tm}
\nu_{eL} \rightarrow \bar{\nu}_{a R} 
\end{equation}
where $a$ denotes another neutrino flavour, either $\mu$ or $\tau$.

In this paper we focus on alternative mechanisms to explain the
deficit of solar neutrinos via the conversion to electron
anti-neutrinos.  The idea is that, even though the nuclear reactions
that occur in a normal star like our Sun do not produce directly
right-handed active neutrinos ($\bar{\nu}_a$) these may be produced
by combining the above transition in \eq{tm} with the standard 
chirality-preserving MSW conversions \cite{MSW} 
\begin{equation}
\label{msw}
\nu_{eL} \rightarrow \nu_{\mu L} 
\end{equation}
through cascade conversions like
\begin{equation}
\label{cascade}
\nu_{eL} \rightarrow \bar{\nu}_{\mu R} \rightarrow \bar{\nu}_{eR}
\qquad
\nu_{eL} \rightarrow \nu_{\mu L} \rightarrow \bar{\nu}_{eR}
\end{equation}

These conversions arise as a result of the interplay of two types of
mixing \cite{akhpetsmi}: one of them, matter-induced flavour mixing,
leads to MSW resonant conversions which preserve the lepton number
$L$, whereas the other is generated by the interaction of a neutrino
transition magnetic moment \cite{LAM,akhmedov91} with the solar
magnetic field, and violates the $L$ symmetry by two units ($\Delta L
= \pm 2$). This $L$-violation is an explicit signature of the Majorana
nature of the neutrino.  The decay of solar neutrinos into a massless
(pseudo)-scalar majoron $\nu \rightarrow \nu' J$ \cite{decay1,decay2},
is another process which violates lepton number \footnote{ However the
solar neutrino matter-induced decay seems marginally possible as a
solution to the solar neutrino problem \cite{majoron}.}.

There are however stringent bounds on the presence of solar electron
anti-neutrinos in the high energy region ($^8$B). These would interact
within the detector through the process $\bar{\nu}_e + p \rightarrow n
+e^+$.  This process, which has an energy threshold of $E_\nu = m_n
-m_p +m_e \simeq 1.8$ MeV, has not been found to occur in the
Kamiokande experiment \cite{suzuki,barbieri91,raghavan91}, nor
in the very recent data from Super-Kamiokande \cite{fiorentini97}.
Also the results from the liquid scintillation detector (LSD) are
negative \cite{aglietta96}.  However, as we show in section 4, the
co-existence of a suppressed production of high-energy $\bar{\nu}_e$'s
and a sizeable flux of anti-neutrinos at energies below $1.8$ MeV can
be easily understood theoretically.  This happens, for example, for
the specific scenario presented in ref. \cite{akhpetsmi}.

In this paper we propose to probe for the possible existence of
$L$-violating processes in the solar interior that can produce an
anti-neutrino component in the neutrino flux. We consider
neutrino-electron scattering in future underground solar neutrino
experiments in the low-energy region, below the threshold for
$\bar{\nu}_e + p \rightarrow n +e^+$, such as is the case for $pp$ or
$^7$Be neutrinos. These should be measured, respectively, in future
experiments such as HELLAZ and BOREXINO, which will have energy
thresholds \cite{hellaz,borexino}
\footnote{For BOREXINO an energy threshold of $T_{th} \simeq 25$ keV 
is expected, but $\epsilon(\omega) \neq 1$ for this region.}
$$
T_{Th} (\mbox{HELLAZ}) = 100 ~\mbox{keV}, \qquad
T_{Th} (\mbox{BOREXINO}) = 250 ~\mbox{keV}~.
$$
We show that neutrino conversions within the Sun can result
in partial polarization of the initial fluxes, in such a way
as to produce a sizeable $\bar{\nu}_e$ component, while being
unobservable at the Kamiokande and Super-Kamiokande experiments.

\section{The Cross-Sections}

The complete expression for the differential cross section of the weak
process $\nu e \rightarrow \nu e$, as a function of the electron
recoil energy $T$, in the massless neutrino limit, can be written as
\cite{semikoz},

\begin{eqnarray}
\label{totdsigma}
 \frac{d \sigma}{dT} (\omega,T)= && \frac{2 G_F^2 m_e}{\pi} \{
P_e \left [ g_{eL}^2 + g_R^2 \left (1- \frac{T}{\omega}\right )^2 -
g_{eL} g_R \frac{m_e T}{\omega^2}\right ] + \nonumber \\
&& +P_{\bar{e}}\left [ g_R^2 + g_{eL}^2 \left (1- \frac{T}{\omega}\right )^2 -
g_{eL} g_R \frac{m_e T}{\omega^2}\right ] + \nonumber \\
&& +P_a \left [ g_{a L}^2 + g_R^2 \left (1- \frac{T}{\omega}\right )^2 -
g_{a L} g_R \frac{m_e T}{\omega^2}\right ] + \nonumber \\
&&  +P_{\bar{a}}
\left [ g_R^2 + g_{a L}^2 \left (1- \frac{T}{\omega}\right )^2 -
g_{a L} g_R \frac{m_e T}{\omega^2}\right ] \}
\end{eqnarray}
where $g_{eL} = \sin^2 \theta_W+0.5$ , $g_{a L} = \sin^2 \theta_W-0.5$
($a=\mu,\tau$) and $g_R = \sin^2 \theta_W$ are the weak couplings of
the Standard Model, and $\omega$ is the energy of the incoming
neutrino. The different rows in this equation correspond to the
contributions of electron neutrinos, electron anti-neutrinos, muon/tau
neutrinos and muon/tau anti-neutrinos, respectively.

The parameter $P_e$ in the equation above is the survival probability 
of the initial left-handed electron neutrinos, while $P_{\bar{e}}$,
$P_a$ and $P_{\bar{a}}$ are the appearance probabilities of the other
species, that may arise in the Sun as a result of the processes
$\nu_{eL} \to \bar{\nu}_{eR}$, $\nu_{eL}\to \nu_{aL}$ or
$\nu_{eL} \to \bar{\nu}_{aR}$, respectively. These parameters
obey the unitarity condition
\begin{equation}
P_e(\omega) + P_{\bar{e}}(\omega) + 
P_a(\omega) + P_{\bar{a}}(\omega) = 1~,
\label{unitarity}
\end{equation}
In general they are obtained from the complete $4\times 4$ evolution 
Hamiltonian describing the evolution of the neutrino system
\cite{BFD} after taking into account the effects of matter
\cite{LAM}. They depend, in general, on the neutrino energy $\omega$,
on the solar magnetic field through the parameter $\mu_\nu B_\perp$
and on the neutrino mixing parameters $\Delta m^2$, $\sin ^22\theta$.

In the $L$-violating processes like the conversions in \eq{cascade},
one has in general all four contributions shown in \eq{totdsigma}. In
contrast, in the case where lepton number is conserved (like in MSW
conversions), the solar neutrino flux will consist of {\em neutrinos},
so only the first and third rows in \eq{totdsigma} contribute. For
$\nu_e \rightarrow \nu_{\mu,\tau}$ (active-active conversions) one has
the contribution of both terms, since then $P_a = 1 - P_e$, while only
the terms proportional to $P_e$ are present in the case of $\nu_e
\rightarrow \nu_s$ (active-sterile) conversions, as the detector is
{\em blind} to sterile neutrinos. It follows that the differential
cross section will be {\em different} in the case where electron
neutrinos from the Sun get converted to electron anti-neutrinos.  The
question one should answer is the following: is it possible to measure
this difference in the present or in future underground neutrino
experiments?

\section{BOREXINO and HELLAZ}

The relevant quantity to be measured in neutrino scattering
experiments is the energy spectrum of events, namely
\begin{equation}
\label{spectrum}
\frac{dN_\nu}{dT} = N_e \sum_{i} \phi_{0i} 
\int^{\omega_{max}}_{\omega_{min}(T)}
d \omega~ \lambda_i(\omega) \epsilon(\omega)
\langle \frac{d \sigma}{dT} (\omega,T) \rangle \:  \:, T > T_{th}
\end{equation}
where $d \sigma/dT$ is given in \eq{totdsigma}, $\epsilon(\omega)$ is
the efficiency of the detector (which we take as unity for energies
above the threshold, for simplicity), and $N_e$ is the number of
electrons in the fiducial volume of the detector.  The sum in the
above equation is done over the solar neutrino spectrum, where $i$
corresponds to the different reactions $i= pp$, $^7$Be, $pep$, $^8$B
$\ldots$, characterized by a differential spectrum $\lambda_i(\omega)$
and an integral flux $\phi_{0i}$.  The lower limit for the neutrino
energy is
$$
\omega_{min} (T) = \frac{T}{2} \left ( 1 + \sqrt{1
+\frac{2m_e}{T}} \right )~.
$$
while the upper limit $\omega_{max}$ corresponds to the maximum
neutrino energy, taken, as $\lambda_i(\omega)$, from the Standard
Solar Model \cite{SSM}. For neutrinos coming from two-body reactions,
like $^7$Be or $pep$ neutrinos, one has $\lambda_i(\omega) =
\delta(\omega - \omega_i)$.

In order to take into account the finite resolution in the measured 
electron recoil energy, we perform a Gaussian average of the cross
section, indicated by $\langle \ldots \rangle$ in \eq{spectrum},
\begin{equation}
\langle \frac{d\sigma}{dT}(\omega, T) \rangle = 
\frac{\int_{0}^{T_{max}} dT' ~\exp 
\Bigl [- \frac{(T' - T)^2}{2\Delta^2_{T'}}\Bigr ]\frac{d\sigma}
{dT'}(\omega, T')}
{\int_{0}^{T_{max}} dT' ~\exp 
\Bigl [- \frac{(T' - T)^2}{2\Delta^2_{T'}}\Bigr ]}~.
\label{gauss}
\end{equation}
Here $T'$ is the {\em true} recoil electron energy in the
cross-section \eq{totdsigma}, and $T$ the measured recoil energy. The
electron recoil energy obeys the following kinematical inequality
$$
0 \leq T' \leq T_{max} = \frac{2 \omega^2}{m_e + 2 \omega}~,
$$

The finite
energy resolution of the experiments has been introduced in the
parameter $\Delta_T \equiv \Delta T$.  For the corresponding
experiments in the energy region below $1$ MeV, one has the following
electron recoil energy resolutions.

The liquid scintillator in BOREXINO is expected to observe
approximately 300 photoelectrons ($phe$) per MeV of deposited electron
recoil energy. This gives an estimate of the energy resolution of the
order \cite{borexino}
\begin{equation}
\label{deltatborex}
\frac{\Delta T}{T} (\mbox{BOREXINO}) \simeq 
\frac{1}{\sqrt{N_{phe}}} \simeq \frac{0.058}{\sqrt{T/\mbox{MeV}}} ~,
\end{equation}
This corresponds to 12\% for the threshold ($T_{th} \simeq 0.25$ MeV)
and 7\% for the maximum recoil energy for $^7$Be neutrinos, 
$T_{max} \simeq 0.663$ MeV.

In HELLAZ the Multi-Wire-Chamber (MWC) counts the secondary electrons
produced by the initial one in helium. It is expected to count 2500
electrons at the threshold recoil energy, $T_{th} \simeq 0.1$ MeV, so
in that case \cite{hellaz}
\begin{equation}
\label{deltathellaz}
\frac{\Delta T}{T} (\mbox{HELLAZ}) \simeq 
\frac{1}{\sqrt{N_e}} \simeq \frac{0.02}{\sqrt{T/T_{th}}} ~,
\end{equation}
or an energy resolution of the order 1.5\% for the maximum for $pp$
neutrinos $T_{max} \simeq 0.26$ MeV.

First let us focus in the simple case where the parameters $P_i$ do
not depend on the neutrino energy.  Since the scattering of $pp$
neutrinos can not lead to electron recoil energies above $0.26$ MeV
approximately, we can consider separately the detection of $pp$
neutrinos in HELLAZ \footnote{HELLAZ is also intended to measure the
contribution from $^7$Be neutrinos. However it is uncertain how they
will separate the contribution of $pp$ neutrinos for energies close to
$T^{pp}_{max} \simeq 0.26$ MeV} for the region $0.1$ MeV $<T<0.26$ MeV
and the corresponding of $^7$Be neutrinos in BOREXINO for $0.26$ MeV
$<T<0.663$ MeV.

\begin{figure}
\centerline{\protect\hbox{\psfig{file=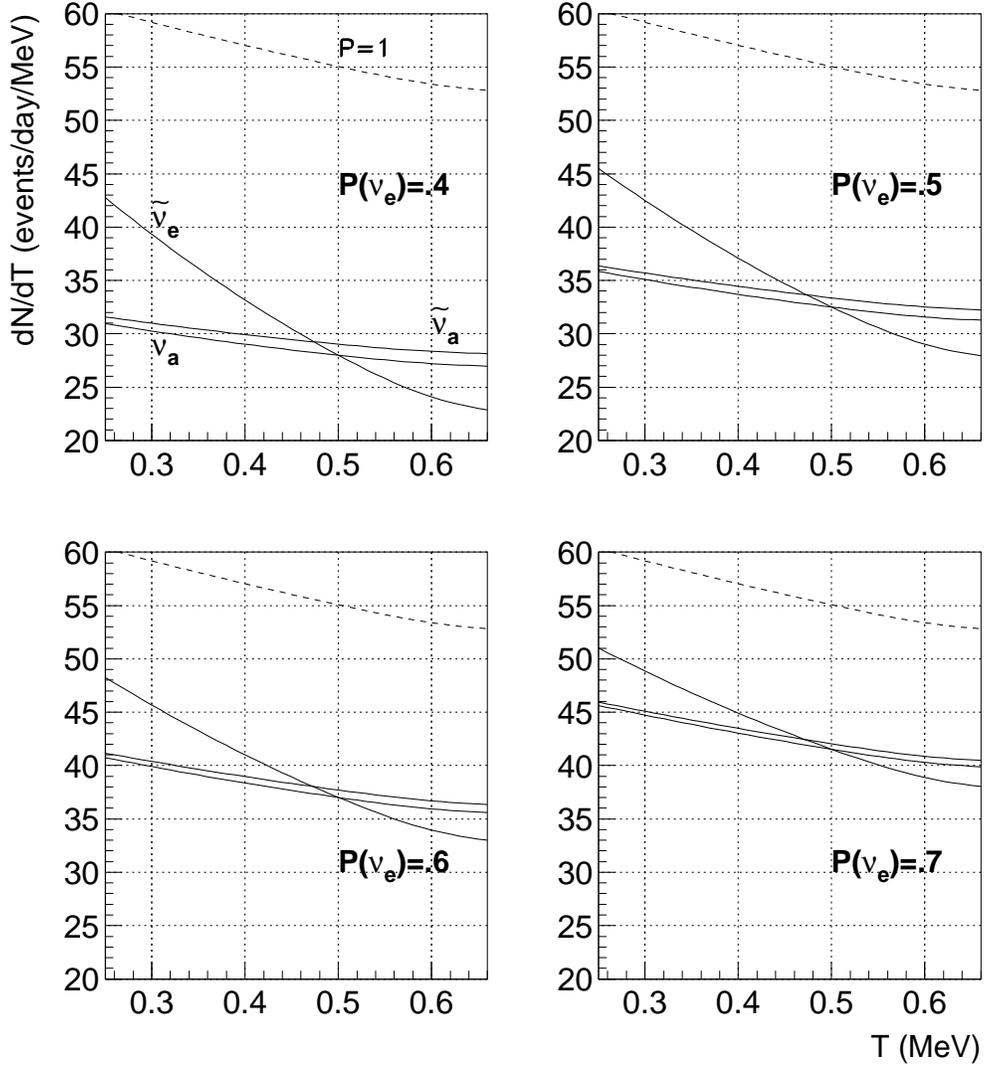,height=14cm}}}
\caption{Energy spectrum of events corresponding to $^7$Be solar
neutrinos for the BOREXINO experiment. Different cases are shown,
where the solar neutrino flux consists of: only electron neutrinos
(label $P_e=1$), electron neutrinos and electron anti-neutrinos
($\bar{\nu}_e$), electron and muon/tau neutrinos ($\nu_a$) and finally
electron and muon/tau anti-neutrinos ($\bar{\nu}_a$).  The electron
neutrino survival probability $P_e$ (except for the upper line) takes
the value as indicated.}
\label{fig1}
\end{figure}

\begin{figure}
\centerline{\protect\hbox{\psfig{file=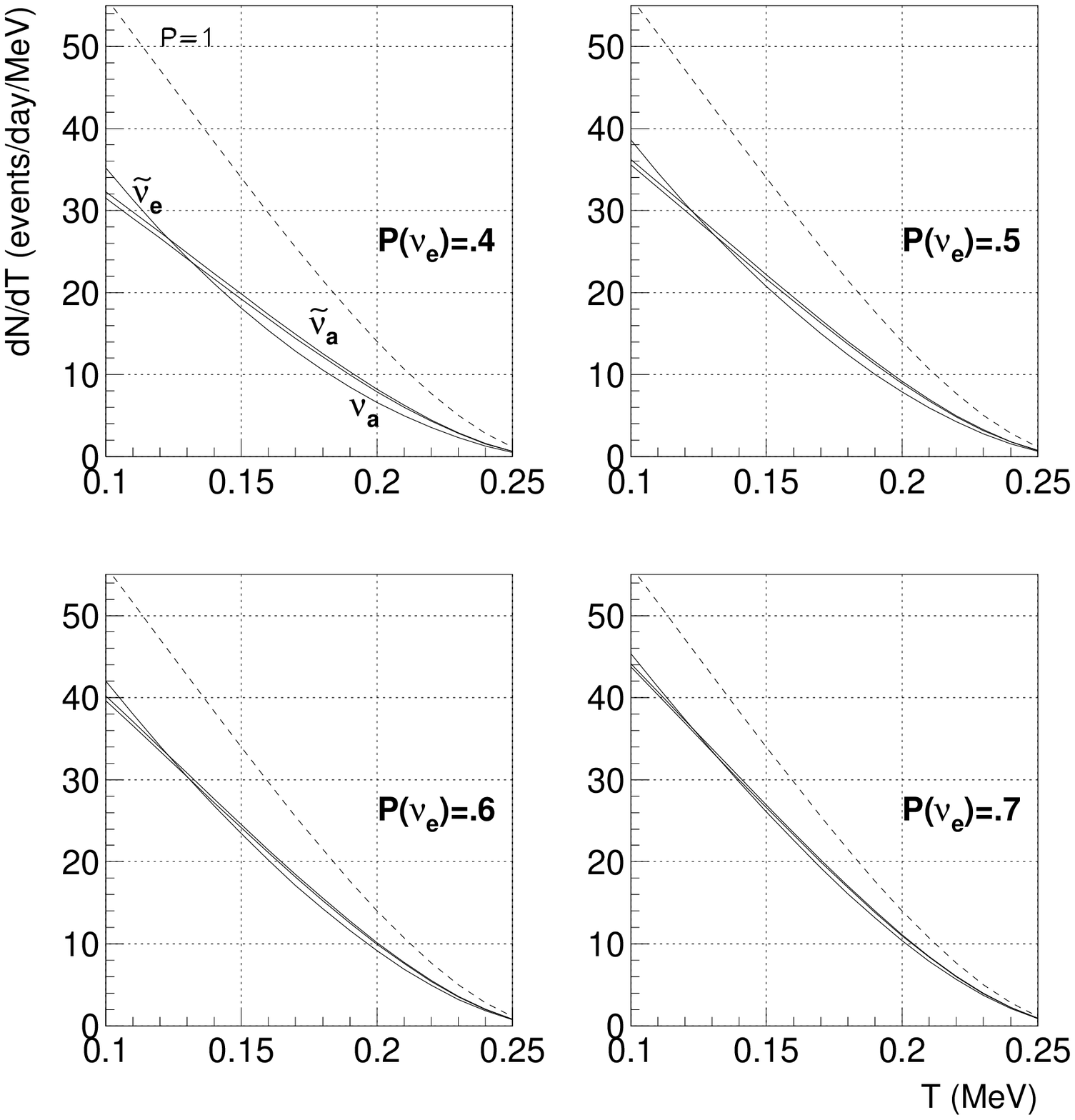,height=14cm}}}
\caption{Same as figure 1 for $pp$ neutrinos and the HELLAZ experiment.}
\label{fig2}
\end{figure}

We have calculated the averaged energy spectrum of events from
\eq{spectrum} for the two experiments. Our results are shown in
figures \ref{fig1} and \ref{fig2} for BOREXINO and HELLAZ,
respectively.  The upper line in the different figures corresponds to
the case where one has no neutrino conversions, so $P_e=1$. When
electron anti-neutrinos are present in the solar flux the results are
the lines labelled with $\bar{\nu}_e$, calculated for the indicated
value of $P_e$ and $P_{\bar{a}}=0.05$ (since for the cascade scenario
in \eq{cascade} one needs at least a small amount of $\bar{\nu}_a$).
The cases of $\nu_e \rightarrow \nu_{\mu,\tau}$ and $\nu_e \rightarrow
\bar{\nu}_{\mu,\tau}$ are the lower lines with labels $\nu_a$ and
$\bar{\nu}_a$, respectively.


One can see from figure \ref{fig1} that it is possible to distinguish
the case with $\bar{\nu}_e$ considering the behaviour of the cross
section for low energies. It is the {\em slope} of the measured
spectrum the key for recognizing the presence of electron
anti-neutrinos in the solar neutrino flux, and correspondingly the
presence of $L$-violating processes which can only exist if neutrinos
are Majorana particles.

Comparing figures \ref{fig1} and \ref{fig2}, we conclude that the
measurement of $^7$Be neutrinos in BOREXINO is more efficient for the
discrimination, because in contrast to the $pp$ case (HELLAZ) the
difference in the slope of the process with electron anti-neutrinos with
respect to the other cases appears well above the energy threshold.

On the other hand one can also see from the diagrams in figure \ref{fig1}
that the discrimination is possible for a broad range of values
for $P_e$, provided that it is not very close to unity nor to zero.

The shortcoming of the above discussion is of course that physical
parameters $P_i$ do depend on the neutrino energy.  One must calculate
the averaged $\nu-e$ cross section using analytical expressions for
$P_i=P_i(\omega)$ for the different processes that have been addressed
to solve the Solar Neutrino Problem (SNP).  However, since the $^7$Be
neutrinos are mono-energetic, whatever the mechanism that produces the
deficit is, their survival probability will be a constant value
of $P_e(\omega_{Be})$. Therefore one can apply directly the results we
have obtained for constant $P_i$ for the range of electron recoil
energy where the contribution of $^7$Be neutrinos dominates
(approx. from $T_{min} = T_{max}(pp~\nu) \simeq 0.261$ MeV to $T_{max}
\simeq 0.665$ MeV). The solar neutrino flux in this region will be
measured with good accuracy in the forthcoming experiments BOREXINO
and HELLAZ.

\vspace{0.5cm}

Other experimental uncertainties must be incorporated to the
differential spectra obtained from \eq{totdsigma}.  For example, in
the results shown we have neglected an unknown statistical error,
since it will decrease as $\mid \pm \Delta N/N \mid \sim t^{-1/2}$.
Thus, after enough running time in the experiment, the statistical
error may be assumed to be less than the systematic error.  Moreover,
for the BOREXINO experiment one expects a small internal background in
the low-energy window, $0.25 \leq T/$MeV$\leq 0.8$. The main
contaminators, such as $^{14}$C and $^{246}$Pa, will be well
discriminated in the liquid scintillator (see fig. 15 in ref.
\cite{borexino}).  External background is estimated to be less than
$0.1$ events per day, mainly from muons of cosmic rays, therefore
expected to be negligible. Finally, our assumption of constant $P_i$
requires that the value of the solar magnetic field is fixed for a
long period.

\section{Kamiokande and Super-Kamiokande Limits}

In this section we show that the conversion of solar neutrinos to
electron anti-neutrinos can be suppressed in the high-energy region of
$^8$B neutrinos, while being sizeable for neutrinos with energies
below $1$ MeV. 

The differential spectrum of electron anti-neutrinos in the $\omega
\gg 1$ MeV region, $\lambda_{\bar{\nu}}(\omega)$, is the corresponding
one of $^8$B solar neutrinos distorted due to multiplication by the
conversion probability
\begin{equation}
\label{distorted}
\lambda_{\bar{\nu}}(\omega) = \lambda_\nu^B(\omega)
P_{\nu_{eL}\to \bar{\nu}_{eR}}(\omega)
\end{equation}
We choose as a particular model the one presented in reference 
\cite{akhpetsmi}. In this model the $\nu_{eL} \rightarrow
\bar{\nu}_{eR}$ conversions occur in a twisting magnetic field 
\cite{as} in the triple resonance case with a probability given as
\begin{equation}
\label{APS}
P_{\bar{e}}(\omega) \simeq A(\omega) \sin^4 \frac {({\cal F}t)}{2}
\end{equation}
where the oscillation depth $A$ and the oscillation frequency 
${\cal F}$ take the form
\begin{equation}
\label{oscidepth}
A(\omega) = \frac{4(\delta \sin 2\theta)^2 (\mu B_\perp)^2}
{[(\delta \sin 2\theta)^2 + (\mu B_\perp)^2]^2} \leq 1
\end{equation}
\begin{equation}
\label{oscifreq}
{\cal F} = \sqrt{(\delta \sin 2\theta)^2+ (\mu B_\perp)^2}
\simeq  \frac{\mu}{2 \times 10^{-11}\mu_B} \frac{B_{\perp}}
{10^4 \mbox{G}} \sqrt{ 1 + \left( \frac{\delta \sin 2\theta}
{\mu B_\perp} \right )^2} \times 10^{-15} \mbox{eV}
\end{equation}
These parameters depend on the value of the magnetic field in the
solar convective zone $B_\perp$, the neutrino transition magnetic
moment $\mu$, the neutrino vacuum mixing angle $\theta$ and, finally,
the neutrino non-degeneracy parameter $\delta=\Delta m^2/4 \omega$.

The oscillation length, $l_{osc}=2\pi /{\cal F}$, must be much less 
than the width of the convective zone $L_{conv} \simeq 3 \times 10^{10}$
 cm $=1.5 \times 10^{15}$ eV$^{-1}$, otherwise
$\sin {\cal F}t/2 \sim {\cal F}t/2 $ and the conversion probability
in eq. (\ref{APS}) is small. In such a case the maximum value of 
$P_{\bar{e}}$ is obtained at the resonance energy, where 
$$
\delta_{res} \sin 2\theta = \frac{\Delta m^2}{4 \omega_{res}} 
\sin 2\theta = \mu B_\perp
$$
Then one can average $\langle \sin^4 ({\cal F}t/2) \rangle = 3/8$
\footnote{Note that here we correct eq. 7 of ref. \cite{semikoz}
in which $\sin^2$ should be substituted for the averaged propagation
factor $\langle \sin^4({\cal F}t)/2\rangle= 3/8$.}, and this is the
maximum value of the conversion probability.

The resonance can lie in the energy region below $1$ MeV, provided
that the neutrino parameters have reasonable values. In such a case
the conversion probability is small for energies $\omega \gg 1$ MeV.
For instance, if the resonant energy coincides with the neutrino
$^7$Be line, $\omega_{res} \simeq 0.862$ MeV, one can estimate that
the conversion probability for neutrino energies above $8$ MeV is
\begin{equation}
\label{estimate}
P_{\bar{\nu}_e}(8 ~\mbox{MeV}) \simeq \frac{3}{2} 
\left (\frac{\delta(8 ~\mbox{MeV})}{\delta(\omega_{res})}\right )^2
= \frac{3}{2} \left (\frac{0.862 ~\mbox{MeV}}{8 ~\mbox{MeV}}\right )^2
\simeq 0.015~.
\end{equation}
i.e, only a few percent of the initial electron neutrinos in the high
energy range convert to electron anti-neutrinos.

Bounds on the solar $\bar{\nu}_e$ flux were obtained from the analysis
of the isotropic background in the Kamiokande \cite{barbieri91} and
Super-Kamiokande \cite{fiorentini97} experiments.  A similar bound has
been derived from the analysis of the experimental data obtained on
the liquid scintillation detector (LSD) \cite{aglietta96}.

Assuming that the anti-neutrino spectrum has the same shape as that
characterizing $^8$B solar neutrinos, one has the following bound on
the anti-neutrino flux \cite{fiorentini97}
\begin{equation}
\label{antiflux1}
\Phi_{\bar{\nu}} (\omega_{\bar{\nu}}> E_0 = 8.3~\mbox{MeV} ) <
6\cdot 10^4~ \mbox{cm}^{-2}\mbox{sec}^{-1}
\end{equation}
This bound sets an upper limit to the presence of electron anti-neutrino
in the $^8$B region, since it must be less than 3.5\% of the solar
neutrino flux predicted by the Standard Solar Model in the corresponding
energy range. 

\begin{figure}
\centerline{\protect\hbox{\psfig{file=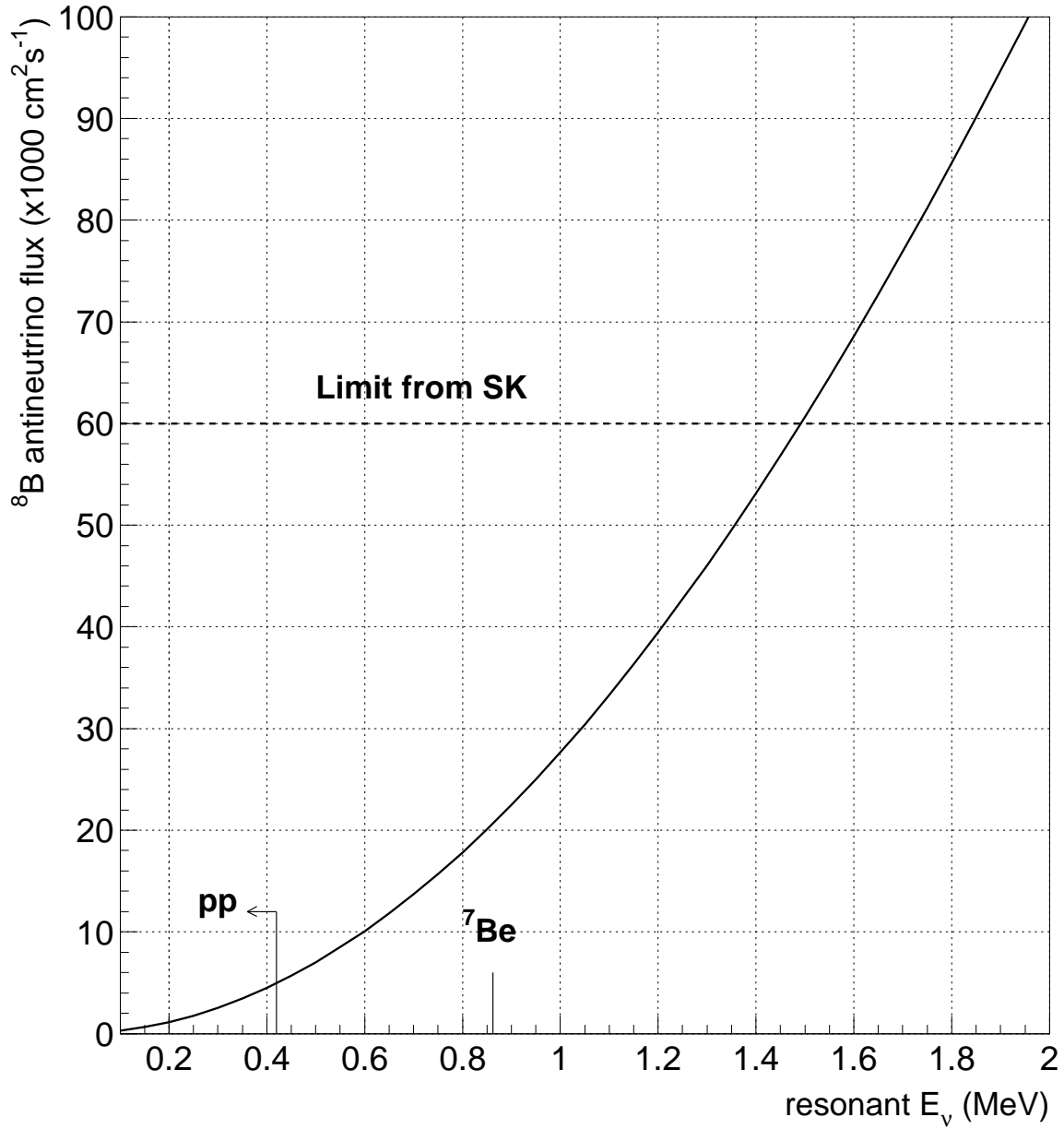,width=0.95\textwidth}}}
\caption{Flux of electron anti-neutrinos in the $^8$B region according to
the conversion mechanism of ref. \protect\cite{akhpetsmi}. We have
fixed the value of $\Delta m^2 \sin 2\theta$ at $10^{-8}$ eV$^2$ and
varied $\mu B_\perp$ so as to have the resonant conversion energy in
the region below $2$ MeV. The horizontal line corresponds to the limit
derived from Super-Kamiokande.}
\label{fig3}
\end{figure}
We have calculated the expected flux of high energy electron
anti-neutrinos in the case that the conversion occurs through
eq. (\ref{APS}). The product of neutrino parameters $\Delta m^2 \sin
2\theta $ has been fixed at $10^{-8}$ eV$^2$, whereas the product $\mu
B_\perp$ has been varied so as to have the resonant conversion energy
in the region below $2$ MeV, relevant for $pp$ or $^7$Be
neutrinos. Since no positron signal from inverse $\beta$ decay has
been observed in (Super)Kamiokande, its contribution must lie below
the flat background. We follow the analysis of reference
\cite{barbieri91} and its updated version for the Super-Kamiokande data
\cite{fiorentini97}.

Using (\ref{APS}) as the conversion probability, the flux of
$\bar{\nu}_e$ will be
\begin{equation}
\label{antiflux2}
\Phi_\nu^{B,SSM}(\omega_\nu > E_0=8.3 ~\mbox{MeV}) \times
\frac{\int^{\infty}_{E_0}d\omega ~ \lambda_\nu^B(\omega) 
\sigma_{\bar{\nu}}(\omega) P_{\bar{\nu}_e}(\omega,\omega)}
{\int^{\infty}_{E_0}d\omega ~ \lambda_\nu^B(\omega) 
\sigma_{\bar{\nu}}(\omega) }
\end{equation}

The above expression is plotted in figure (\ref{fig3}) as a function
of the neutrino energy where the resonance takes place. The bound
derived in \cite{fiorentini97} corresponds to the horizontal line. One
can see from this example that the anti-neutrino flux would be {\em
hidden} in the background and therefore unobservable in
Super-Kamiokande if $\omega_{res}$ lies in the region of $pp$ or
$^7$Be solar neutrinos, relevant for the HELLAZ or BOREXINO
experiments, as we discussed in the previous section.

\section{Conclusions}

In this paper we have argued that the observation of electron
low-energy anti-neutrinos from the Sun could lead to the conclusion
that the neutrinos are Majorana particles, without conflicting present
Kamiokande or Super-Kamiokande data. It is important to emphasize that
in the conversions we assume, either given by \eq{cascade} or
presumably caused by \ne decay, the violation of total lepton number
is not produced {\em directly} by the a Majorana neutrino mass. This
is in contrast to the case of laboratory experiments, where the
differences between Dirac and Majorana neutrinos can only arise via a
neutrino mass insertion and are therefore helicity-suppressed
\cite{1666}. This is because the neutrino beams produced in laboratory
are {\em fully-polarized}. This applies to neutrinos produced by the
weak decay of mesons from accelerators or in reactor or isotope
neutrino sources.

The Sun, however, can possess a large-scale magnetic field in the
convective zone ($L_{conv}\simeq 3\times 10^{10}$ cm) with a
relatively modest value $B_{\perp}\sim 10^4$ G. This would effectively
cause a neutrino spin-flip if $\mu_{\nu}B_{\perp}L\sim 1$ for
experimentally acceptable values of the neutrino transition magnetic
moment, $\mu_{ij}\sim 10^{-11}\mu_B$.  This way one can obtain a flux
of neutrinos from the Sun which is {\em partially-polarized}, in the
sense that both neutrinos and anti-neutrinos are present.

Notice that in order to have the conversions in \eq{cascade}, one must
require the presence of the Resonant Spin--Flavour Precession (RSFP)
mechanism as an intermediate step.  As recently discussed in a nice
review by Akhmedov \cite{revAkhmedov}, this mechanism is not in
contradiction with the non-observation of time variations in
Kamiokande or GALLEX--SAGE neutrino experiments.

\section*{Acknowledgements}

The authors thank Gianni Fiorentini, Franz von Feilitzsch, Carlo
Giunti, Alexei Smirnov, Stephan Sch\"onert and Tom Ypsilantis for
fruitful discussions.  This work has been supported by DGICYT under
Grants PB95-1077 and SAB95-506 (V.B.S.), by the TMR network grant
ERBFMRXCT960090 and by INTAS grant 96-0659 of the European
Union. S.P. was supported by Conselleria d'Educaci\'o i Ci\`encia of
Generalitat Valenciana. V.B.S. also acknowledges the support of RFFR
through grants 97-02-16501 and 95-02-03724.

\end{document}